\documentclass[aip,rsi,amsmath,amssymb,reprint]{revtex4-1}
\usepackage{graphicx}
\usepackage{dcolumn}
\usepackage{bm}
\usepackage{braket}

\begin{document}

\preprint{AIP/123-QED}

\title{Nonlinear characterisation of a silicon integrated Bragg waveguide filter}

\author{Micol Previde Massara}
\affiliation{Dipartimento di Fisica, Universit\'a degli Studi di  Pavia, via A. Bassi 6, 27100 Pavia, Italy}

\author{Matteo Menotti}
\affiliation{Dipartimento di Fisica, Universit\'a degli Studi di  Pavia, via A. Bassi 6, 27100 Pavia, Italy}

\author{Nicola Bergamasco}
\affiliation{Dipartimento di Fisica, Universit\'a degli Studi di  Pavia, via A. Bassi 6, 27100 Pavia, Italy}

\author{Nicholas C. Harris}
\affiliation{Department of Electrical Engineering and Computer Science, Massachusetts Institute of Technology, 77 Massachusetts Avenue, Cambridge, MA 02139, USA}

\author{Tom Baehr-Jones}
\affiliation{Elenion Technologies, LLC, New York, NY, USA}

\author{Michael Hochberg}
\affiliation{Elenion Technologies, LLC, New York, NY, USA}

\author{Christophe Galland}
\affiliation{Ecole Polytechnique F\'{e}d\'{e}rale de Lausanne, Institute of Physics, EPFL SB IPHYS GR-GA, CH-1015 Lausanne, Switzerland}

\author{Marco Liscidini}
\affiliation{Dipartimento di Fisica, Universit\'a degli Studi di  Pavia, via A. Bassi 6, 27100 Pavia, Italy}

\author{Matteo Galli}
\affiliation{Dipartimento di Fisica, Universit\'a degli Studi di  Pavia, via A. Bassi 6, 27100 Pavia, Italy}

\author{Daniele Bajoni}
\email{bajdan08@unipv.it}
\affiliation{Dipartimento di Ingegneria Industriale e dell'Informazione, Universit\'a degli Studi di Pavia, via Ferrata 1, 27100 Pavia, Italy}

          
\begin{abstract}
Bragg waveguides are promising optical filters for pump suppression in spontaneous Four-Wave Mixing (FWM) photon sources. In this work, we investigate the generation of unwanted photon pairs in the filter itself. We do this by taking advantage of the relation between spontaneous and classical FWM, which allows for the precise characterisation of the nonlinear response of the device. The pair generation rate estimated from the classical measurement is compared with the theoretical value calculated by means of a full quantum model of the filter, which also allows to investigate the spectral properties of the generated pairs. We find a good agreement between theory and experiment, confirming that stimulated FWM is a valuable approach to characterise the nonlinear response of an integrated filter, and that the pairs generated in a Bragg waveguide are not a serious issue for the operation of a fully integrated nonclassical source. 
\end{abstract}

\maketitle

Silicon ridge waveguides and ring resonators have been shown to be very efficient integrated sources of quantum states of light \cite{Azzini2012OL,Grassani:Optica:2015,Silverstone2015,Engin2013,Davanco2012,Caspani_review}. Despite the large field enhancement that can be achieved in these structures, the efficiency of Spontaneous Four-Wave Mixing (SFWM) is relatively low, with the average number of photon pairs generated being typically 9-10 orders of magnitude smaller than that of the pump photons. Thus, the simultaneous integration of the source with the detection stage on the same chip requires the development of an optical filter capable of 100 dB of pump rejection. This is particularly challenging, for all the frequencies of interest are in the same spectral region, with the signal and idler frequencies symmetrically spaced around the pump. Recent progresses in the integration of such a filter in a silicon chip have been reported exploiting three main strategies: coupled ring resonators \cite{Ong2014}, Bragg waveguides \cite{Harris:PRX:2014} and cascaded interferometers \cite{Piekarek:17}. 

These approaches rely on optical elements composed of hundreds of microns of silicon waveguide, which are potential sources of unwanted photon pairs whose spectral and temporal correlations are usually  different from those of the photons emitted by the actual source. For instance, in the case of heralded single-photon sources, these parasitic photons could lower the purity of the heralded single-photon state, thus reducing dramatically the performance of the entire device. The purity of the heralded state is indeed of pivotal importance for most quantum information protocols, in which one envisions a large number of multiplexed integrated sources \cite{Collins:2013eu,Xiong:13,Xiong:2016}  for application in linear optical quantum computation and simulation \cite{Peruzzo2010,Knill2001,Crespi2011,Crespi2013,Spring2013,Tillmann2013,Harris:2017}.

In this work, we investigate the generation rate and spectral correlations of parasitic photon pairs generated by SFWM in a Bragg waveguide (BW). Our experimental approach exploits the connection between spontaneous and stimulated FWM \cite{Liscidini:2013} and is supported by a theoretical quantum model of pair generation in the integrated structure.  

\begin{figure}
\includegraphics[width=\columnwidth]{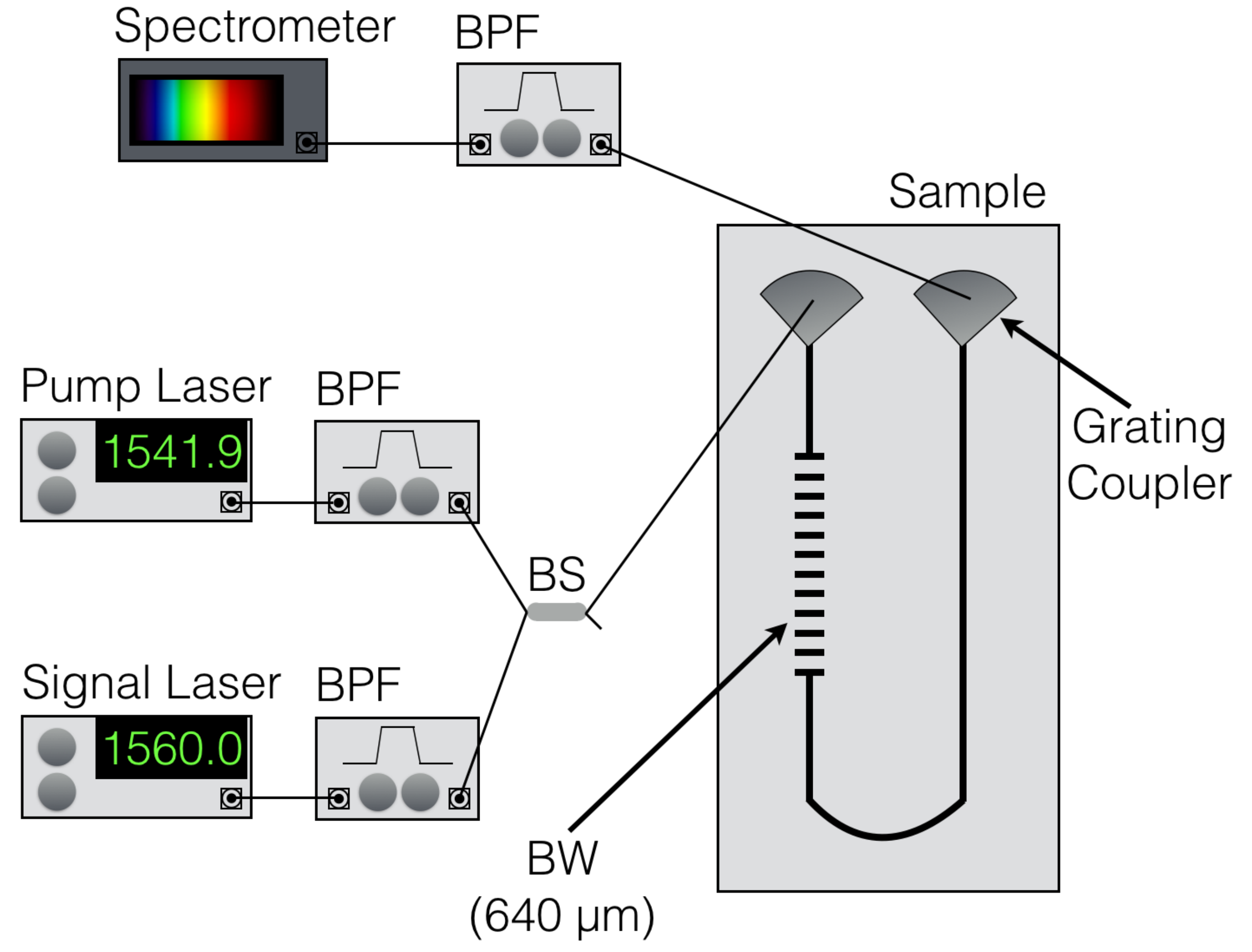} 
\caption{\label{fig:1} Schematic view of the experimental setup used to study four-wave mixing in the Bragg waveguide (BW). BPF stands for band-pass filter and BS for beam splitter. A scheme of the sample used is shown as an inset. The scheme is not in scale. The signal wavelength was kept fixed to 1560 nm, whereas the pump wavelength was scanned from 1541.9 nm to 1550 nm in order to probe the FWM process inside the Bragg waveguide stopband.}
\end{figure}

The sample was fabricated in a CMOS-compatible foundry service (OpSIS \cite{BourzacOpsis2012}) and realized by a 248-nm lithography process on an 8-inches (20.32 cm) silicon-on-insulator wafer. The 220-nm-thick epitaxial silicon layer, with a bulk refractive index $n_{Si}=3.48$ at 1550 nm, is on top of 2 $\mu$m of buried oxide and covered by 2 $\mu$m of oxide cladding (bulk index $n_{SiO_{2}}=1.46$). The ridge waveguide is 500 nm wide and is designed to support a single guided mode in the 1500-1600 nm wavelength range.

\begin{figure}
\includegraphics[width=\columnwidth]{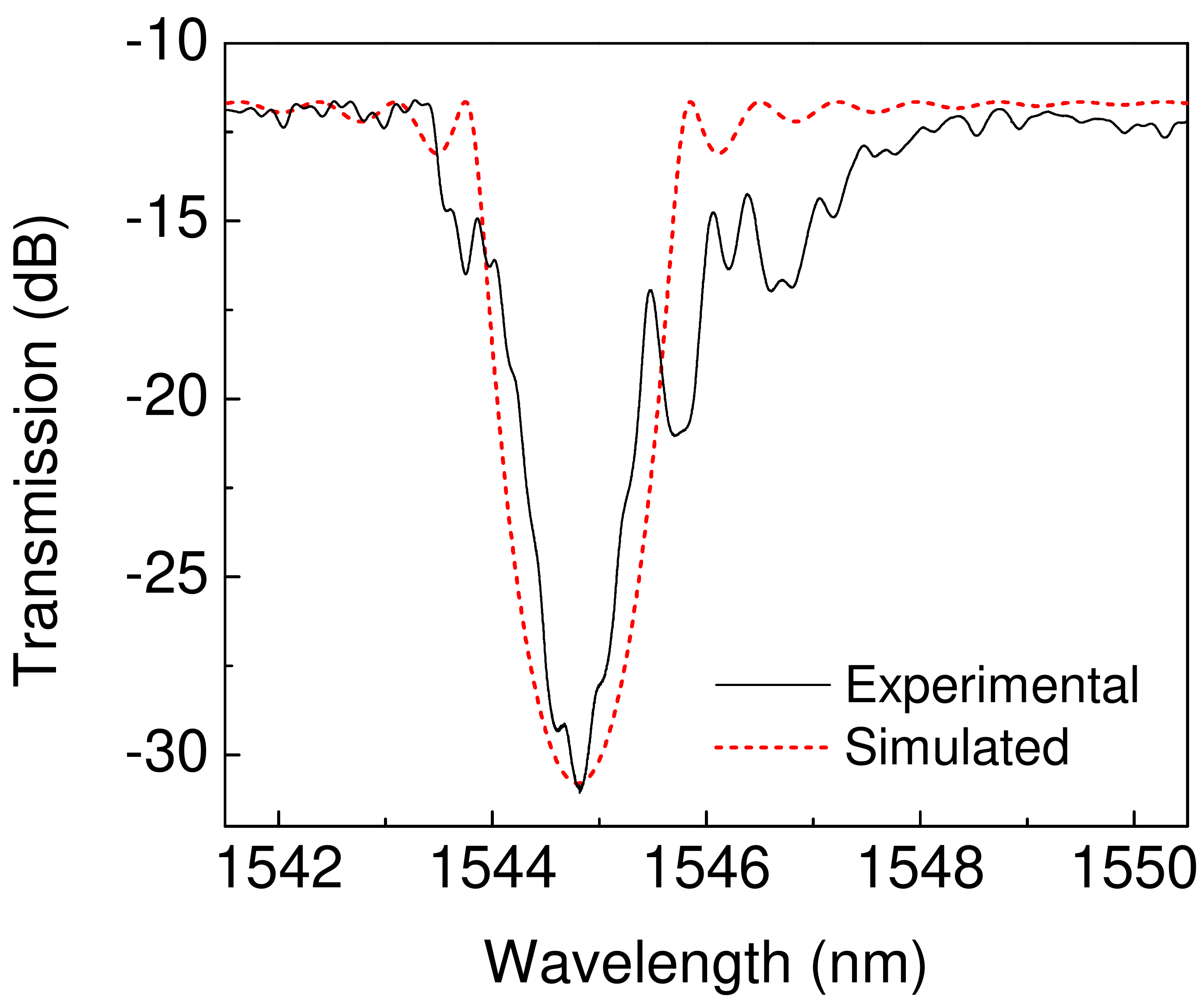} 
\caption{\label{fig:2} Transmission spectrum of the Bragg waveguide around the stopband.}
\vspace{-10pt}
\end{figure}

The BW was designed to achieve high reflectivity in a 1-2 nm wide stopband centered around $\lambda_B=1545$ nm and high transmission outside this range. This was done by periodically shrinking the waveguide width from 500 nm down to 440 nm, with period $\Lambda = \lambda_0 / 2n_{eff} = 320$ nm, duty cycle of 50\% and number of periods $N=2000$. A schematic of the sample is shown as an inset in Fig. \ref{fig:1}. 

In Fig. \ref{fig:2}, we show a high resolution transmission spectrum (2 pm) of the BW, obtained by using a tunable CW infrared laser (Santec TSL-510) and an InGaAs detector (Newport 918D-IG-OD3). The coupling losses at the central wavelength are estimated to be -5 dB for each grating coupler \cite{Harris:PRX:2014}, and the total  insertion losses of the sample are measured to be about -11 dB. The total length of the waveguide is 1.6 mm, of which the BW takes 640 $\mu$m. The spectrum shows a strong reflection around $\lambda = 1544.8$ nm, with a rejection of about 20 dB at the center of the stopband and more than 10 dB rejection over a 1 nm bandwidth.
In Fig. \ref{fig:2} we also report a transmission spectrum calculated using a transfer matrix method and assuming an effective refractive index of the 440 nm-wide waveguide $n_{eff}=2.414$, with an effective refractive index contrast $\Delta n_{eff}=3.4985\cdot 10^{-3}$ with respect to the 500 nm-wide waveguide, which can be calculated using the relation \cite{YarivYeh}:
\begin{equation}\label{Eq:Stopband}
4\Big(1+\frac{\Delta n_{eff}}{n_{eff}}\Big)^{-2N}=10^{-\alpha(dB)/10},
\end{equation}
where $\alpha$ is the rejection in dB.

\begin{figure}
\includegraphics[width=\columnwidth]{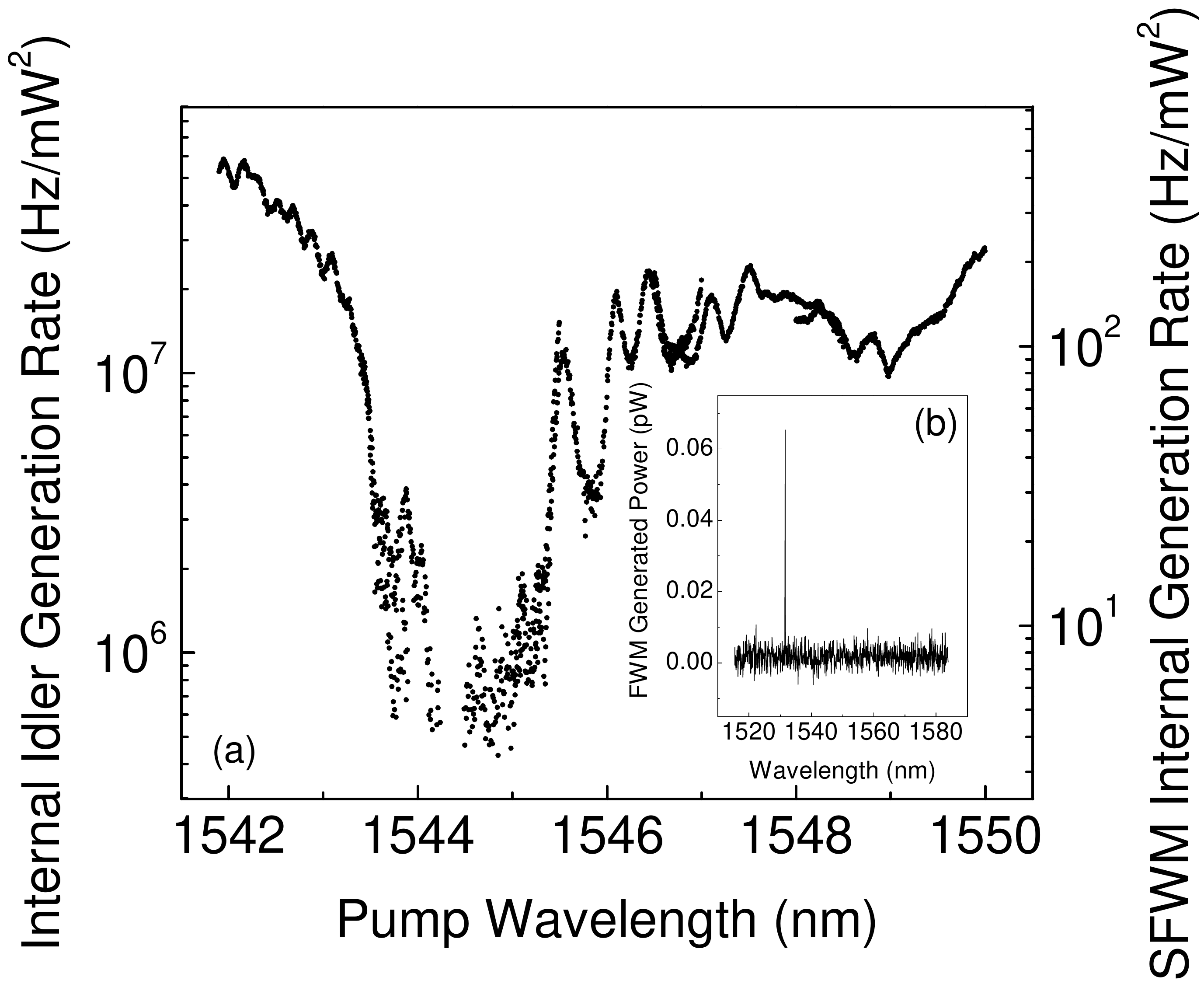}  
\caption{\label{fig:3} (a) Idler generation rate per mW\textsuperscript{2} inside the chip (left y-axis) as a function of the pump wavelength. On the right y-axis we report also the inferred generation rate per mW\textsuperscript{2} for the SFWM process. (b) One spectrum of the idler photons for $\lambda_p = 1545.6$ nm (acquisition time of 0.3 seconds).}
\end{figure}

Stimulated FWM is experimentally investigated by coupling two CW infrared lasers at the pump and the signal frequencies. A schematic layout of the experimental apparatus is shown in Fig. \ref{fig:1}. With the signal laser fixed at 1560 nm, we let the pump wavelength scan the range from 1541.9 nm to 1550 nm, to probe the FWM process across the BW stopband. Before being injected into the chip, the pump and the signal lasers are spectrally filtered by means of a 50 dB band-pass filter, to clean out the amplified spontaneous emission. Next, they are combined on a 90:10 fiber Beam Splitter (BS) and injected into the sample through a fiber array. The optical powers coupled to the sample are 1.29 mW and 1.23 mW for the pump and signal, respectively. Light coming out of the sample is filtered through band-pass filters (Semrock single-band band-pass filters centred at 1550 nm) to suppress the residual pump and signal fields. Finally, we collect the idler output on a spectrometer equipped with a liquid-nitrogen cooled CCD camera. In Fig. \ref{fig:3} we report the estimated internal generation rate of idler photons as a function of the pump wavelength, per mW\textsuperscript{2} of coupled pump power, obtained after calibrating the CCD response against a high-sensitivity power meter and compensating for the transmission of the output filters. The strong suppression of FWM corresponds to the BW stopband.

In a CW pumping scheme, the SFWM process associated with the stimulated process is too faint to be observed experimentally in a BW. However, the spontaneous emission rate $P_{i,spont}$ can be directly connected to the stimulated emission rate $P_{i,stim}$ via the relation \cite{Liscidini:2013}:
\begin{equation}
\label{eq:GENERATION_RATE_SPONT_FWM}
{P_{i,spont} = \hbar\omega_i \Delta\omega \frac{P_{i,stim}}{P_{s}} },
\end{equation}
where $P_{s}$ is the coupled signal power and $\Delta\omega$ is the emission bandwidth. We calculated $P_{i,spont}$ for $\Delta\omega = 2\pi \times 10$ GHz, a typical value for integrated sources, resonant or post-filtered, which corresponds to a reasonable quality factor $Q \approx 20000$. Such sources typically yield generation rates larger than 1 MHz in a 10 GHz bandwidth \cite{Grassani:2016:JSD:scirep,Caspani_review}.

\begin{figure}
\includegraphics[width=\columnwidth]{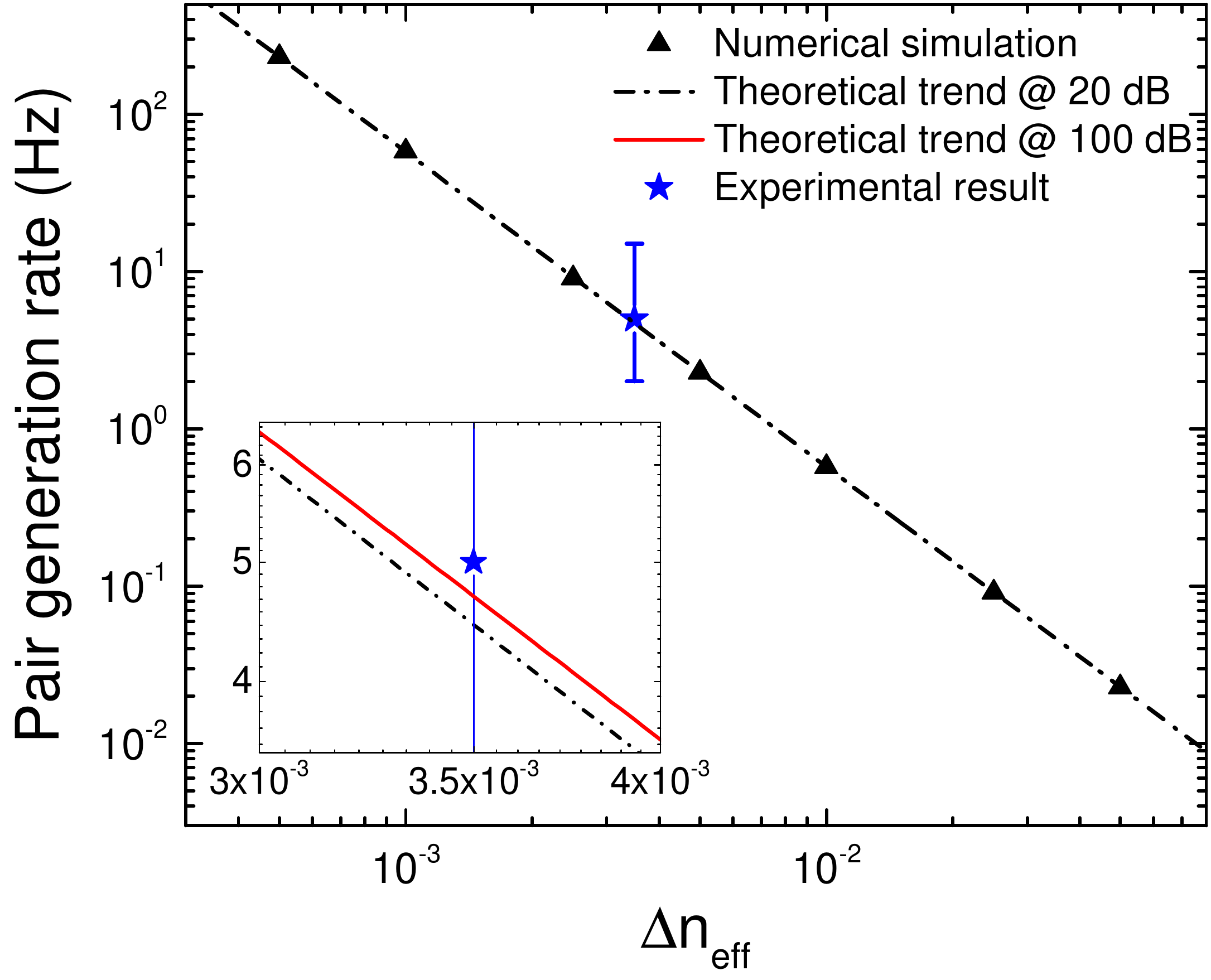}  
\caption{\label{fig:4} Pair generation rate in the Bragg waveguide as a function of the index contrast.}
\vspace{-10pt}
\end{figure}

We report the expected spontaneous emission rate from the BW on the right axis of Fig \ref{fig:3}, where we notice that the average rate of photon pairs generated at the bottom of the stopband is 5 Hz (see also the experimental result in Fig. \ref{fig:4}). Since our BW already provides 20 dB of pump rejection, we expect this figure to be very close to the total generation rate observable in a longer structure as well. This generation rate per mW\textsuperscript{2} is at least 5 orders of magnitude smaller than what efficient silicon integrated sources can produce, thus ruling out any hypothesis of spurious contributions introduced by the filter, even for the most exacting schemes where many sources are multiplexed \cite{Peruzzo2010,Crespi2013,Spring2013,Tillmann2013,Collins:2013eu,Xiong:13,Xiong:2016}.

In Fig \ref{fig:4}, we show the theoretical prediction for the pair generation rate as a function of the refractive index contrast when the BW is modeled using the same parameters adopted for the transmission spectrum, together with the experimental result from our sample.

The results are obtained by means of numerical simulations of the structure based on coupled mode theory \cite{YarivYeh}, the asymptotic-fields method \cite{Liscidini:2012} and the backward Heisenberg picture approach\cite{Yang:2008}. In our computation, we assume a fixed 20 dB extinction rate for the BW and, given a specific effective index contrast, we adjust the number of periods of the BW using Eq. \eqref{Eq:Stopband}.
Since the probability to generate a photon pair by SFWM in the BW is very low, we are working in the undepleted pump approximation, so that the state of the frequency-converted photons is given by
\begin{equation}
\ket{\psi}=\ket{\mathrm{vac}}+\beta\ket{\mathrm{II}}+\cdots,
\end{equation}
where the ellipses refers to higher order terms, which can be neglected in our case, $|\beta|^2$ is the pair generation probability, and $\ket{\mathrm{II}}$ is the normalized two-photon state
\begin{equation}
\ket{\mathrm{II}}=\frac{1}{\sqrt{2}}\int{\ d\omega _1d\omega _2\phi (\omega _1,\omega _2)a^\dagger _{\omega 1}a^\dagger _{\omega 2}\ket{\mathrm{vac}}},
\end{equation}
where $\phi(\omega _1,\omega _2)$ is the biphoton wave function (BWF) and $a^\dagger _{\omega i}$ is the creation operator of a photon with frequency $\omega _i$.
In our simulation, we assume a waveguide nonlinear parameter $\gamma=200$ m\textsuperscript{-1}W\textsuperscript{-1}, which is typical of silicon nanowires \cite{Azzini2012OL}, and modify the effective index of the BW corrugations by $\Delta{n_{eff}}$ over the baseline value $n_{eff}$ of the unperturbed waveguide.
The pump central wavelength 1544.8 nm corresponds to the center of the stopband. We take a 1 ns top-hat temporal profile \cite{Onodera:2016} to guarantee that the SFWM generation rate converges to its CW limit. The theoretical trend reported in Fig. \ref{fig:4} refers to the pair generation rate per 1 mW of coupled pump power when the idler and signal photons are collected in a spectral interval $2\pi\times10$ GHz wide around 1560.05 nm and 1529.94 nm, respectively.
Since the generated idler power is proportional to the square of the waveguide length and the BW length is inversely proportional to the index contrast $\Delta n_{eff}$ at a target extinction ratio, we expect $P_I\propto (\Delta n{eff})^{-2}$, which is well verified by our simulations.
In Fig. \ref{fig:4} we also report the experimental result, which is in good agreement with the theoretical prediction. The figure reports the calculated generation rates for two cases: the measured 20 dB rejection filter, and a 100-dB rejection filter as would be required for a complete pump suppression. Notice that the pair generation rate is almost identical, the difference between the two cases being much smaller than the experimental error bar. Indeed, most of the generation occurs in the first part of the BW, as the pump power decays exponentially within the BW. Therefore, increasing the length would not significantly alter the number of generated pairs.

\begin{figure}
\includegraphics[width=\columnwidth]{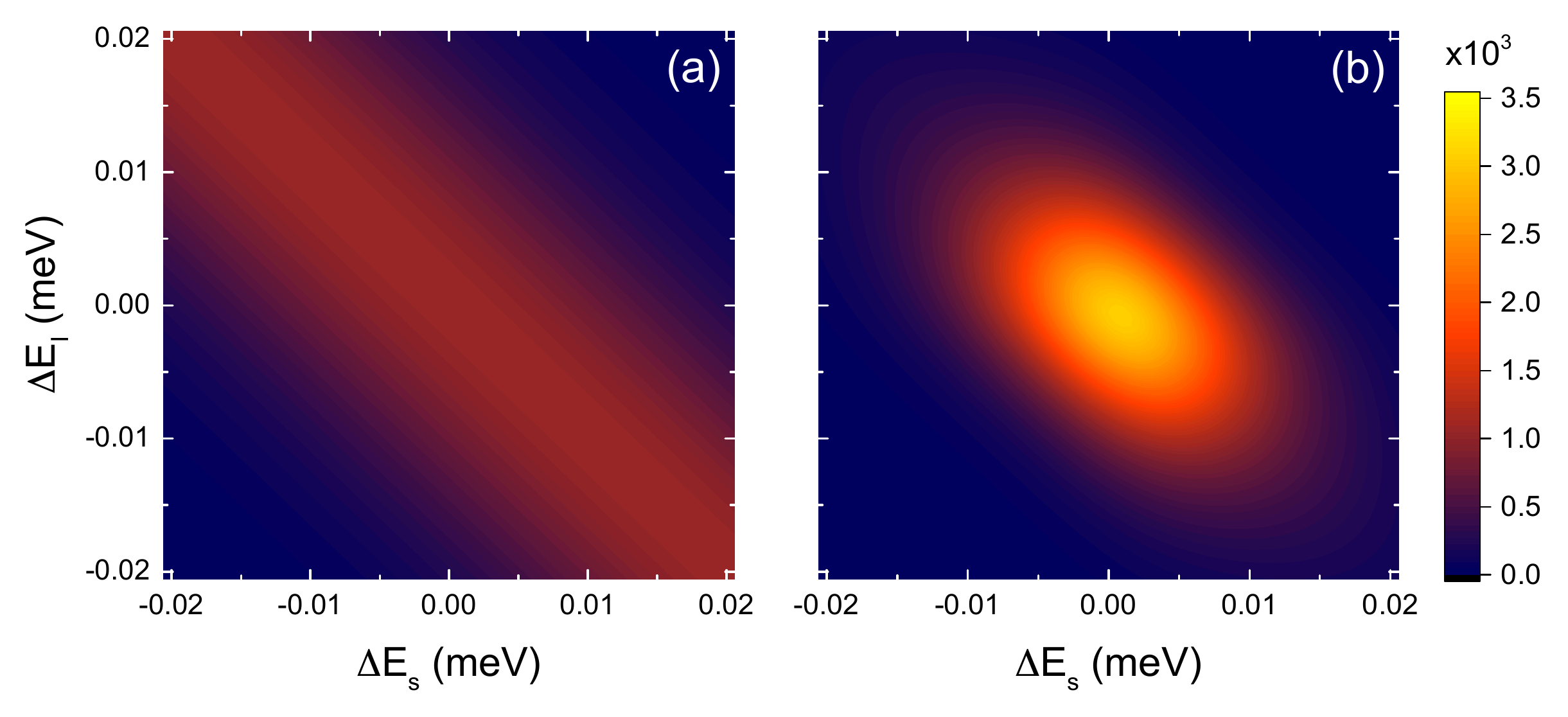}  
\caption{\label{fig:5} Comparison between the JSD of the photon pairs generated by SFWM in (a) a $20\ dB$ Bragg waveguide, and (b) a side-coupled microring resonator.}
\end{figure}

Another relevant aspect of integrated filters is the characterization of the spectral quantum correlations of the photon pairs generated by SFWM in the filter itself.
In Fig. \ref{fig:5} we show the joint spectral density (JSD, the square modulus of the BWF) of the frequency-converted photons generated in the BW, along with that of photon pairs that would be generated by a source of heralded single photon states based on SFWM, represented by an integrated microring resonator. In this scheme, when the pump duration is comparable or shorter than the photon dwelling time in the resonator, one can obtain the generation of nearly uncorrelated photon pairs, a key requirement for heralding single photons in a pure quantum state \cite{Vernon:2017}.

For our comparison, we consider a $15\ \mu$m-radius side-coupled microring resonator, composed of a SOI ridge waveguide. We assume that all the resonances involved in the SFWM process ($\lambda _P=1534.55$ nm, $\lambda _S=1544.27$ nm, and $\lambda _I=1524.94$ nm) have a quality factor $Q=40000$, which corresponds to a dwelling time $\tau_d=33$ ps. Accordingly, we shape the pump pulse with a Gaussian profile with such temporal width. The resulting JSD, reported in Fig. \ref{fig:5}(b), is approximately circular, which is characteristic of nearly uncorrelated photon pairs.
In Fig. \ref{fig:5}(a) we report the JSD of the photons pairs in our filter. In this case, as expected, we observe the generation of highly anti-correlated photons pairs, a typical feature of straight waveguides. In both cases, the waveguide nonlinear parameter, pump power and waveguide dispersion are assumed identical.
In particular, for a 1 mW pump power and a nonlinear waveguide parameter $\gamma=200$ W\textsuperscript{-1}m\textsuperscript{-1}, we calculate a generation probability per pulse in the microring $|\beta_{Ring}|^2=1.0427\cdot 10^{-3}$, which compares favorably to the same figure in the BW $|\beta_{BW}|^2=1.5366\cdot 10^{-11}$.

In conclusion we have nonlinearly characterized a BW intended for pump filtering. These structures are among the most promising for on-chip pump filtering, as they do not require to be actively tuned. Our sample was realized in a silicon photonics fab by using a CMOS compatible process, meaning these filters are widely accessible for general photonics applications. Our results show that BW filters are also ideal structures to provide large pump extinction while avoiding spurious nonlinear processes that may decrease the fidelity of on-chip generated quantum states.

\bigskip
 We acknowledge the University of Pavia Blue Sky Research project number BSR1732907.

\end{document}